%% file: main.tex
\documentclass[manuscript, nonacm]{acmart}

\AtBeginDocument{%
  }

\begin{document}

\title[Human-AI Collaboration with Multi-Agent Generative AI Tools]{Exploring Human-AI Collaboration Using Mental Models of Early Adopters of Multi-Agent Generative AI Tools}

\author{Suchismita Naik}
\email{naik33@purdue.edu}
\orcid{0009-0002-5667-4576}
\affiliation{
    \institution{Purdue University}
    \city{West Lafayette}
    \state{Indiana}
    \country{USA}
}

\author{Austin L. Toombs}
\email{altoombs@iu.edu}
\orcid{10000-0003-3418-0065}
\affiliation{%
  \institution{Indiana University}
  \city{Bloomington}
  \state{Indiana}
  \country{USA}
  \postcode{47408}
}

\author{Amanda Snellinger}
\email{amanda.snellinger@microsoft.com}
\orcid{0009-0006-4194-0535}
\affiliation{
    \institution{Microsoft Research}
    \city{Redmond}
    \state{WA}
    \country{USA}
}

\author{Scott Saponas}
\email{ssaponas@microsoft.com}
\orcid{0000-0002-8806-0125}
\affiliation{
    \institution{Microsoft Research}
    \city{Redmond}
    \state{WA}
    \country{USA}
}

\author{Amanda K. Hall}
\email{amanda.hall@microsoft.com}
\orcid{0000-0001-6151-1814}
\affiliation{
    \institution{Microsoft Research}
    \city{Redmond}
    \state{WA}
    \country{USA}
}

\renewcommand{\shortauthors}{Naik et al.}

\begin{abstract}
  The rise of generative AI tools has transformed human-AI collaboration, particularly in workplace settings. With recent advancements in multi-agent generative AI (Gen AI), technology organizations like Microsoft are adopting these complex tools, redefining AI agents as active collaborators in complex workflows rather than as passive tools. In this study, we investigated how early adopters and developers conceptualize multi-agent Gen AI tools, focusing on how they understand human-AI collaboration mechanisms, general collaboration dynamics, and transparency in the context of AI tools. We conducted semi-structured interviews with \textcolor{black}{13} developers, all early adopters of multi-agent Gen AI technology \textcolor{black}{who work at Microsoft}. Our findings revealed that \textcolor{black}{these} early adopters conceptualize multi-agent systems as ``teams'' of specialized role-based and task-based agents, such as assistants or reviewers, structured similar to human collaboration models and ranging from AI-dominant to AI-assisted, user-controlled interactions. We identified key challenges, including error propagation, unpredictable and unproductive agent loop behavior, and the need for clear communication to mitigate the \textcolor{black}{layered transparency} issues. Early adopters' perspectives \textcolor{black}{about the role of transparency} underscored \textcolor{black}{its importance} as a way to build trust, verify and trace errors, and prevent misuse, errors, and leaks. The insights and design considerations we present contribute to CSCW research about collaborative mechanisms with capabilities ranging from AI-dominant to AI-assisted \textcolor{black}{interactions}, transparency and oversight strategies in human-agent and agent-agent interactions, and \textcolor{black}{how humans make sense of these multi-agent systems as dynamic, role-diverse collaborators which are} customizable \textcolor{black}{for} diverse needs and workflows. We conclude with future research directions that extend CSCW approaches to the design of inter-agent and human mediation interactions.
\end{abstract}

\maketitle

\section{Introduction} 
The integration of multi-agent generative AI (Gen AI) tools in workplace settings represents a pivotal shift in human-AI collaboration for some users, positioning AI as not merely a tool but potentially as an active partner in decision-making and cooperative work. Within the context of human-AI teaming, multi-agent systems signify a new paradigm that further complicates this landscape as we transcend traditional task execution, requiring advanced collaborative capabilities between users and multiple AI agents. This development invites us to rethink the nature of human-AI interaction, as these tools transition from singular functions toward collective, inter-agent dynamics that can augment human tasks, support decision-making, and streamline workflows. Currently, developers and early adopters are shaping the baseline of how multi-agent gen AI systems will be understood and incorporated into existing systems. Therefore, \textcolor{black}{it is important to} document and \textcolor{black}{analyze the metaphors and conceptual models that developers, as early adopters, use to make sense of these tools, including how they determine when and how to use them for collaboration, and what they believe is the capability or challenges of such collaborations.}

As AI systems increasingly \textcolor{black}{adopt} roles traditionally held by human team members, such as collaborators, assistants, or administrators, there is a pressing need to explore the diverse roles AI can perform and how these roles affect work outcomes \cite{garvey_considering_2022, berretta_defining_2023}. Studies by Berretta et al. \cite{berretta_defining_2023}, Dorton \& Harper \cite{dorton_naturalistic_2022}, Wang \& Yin \cite{wang_are_2021}, and Zhang et al. \cite{zhang_i_2024} highlight the importance of understanding the systematic division of labor between humans and AI, as well as the implications of AI's explanatory capabilities on trust and reliance. While attributes like anthropomorphism can enhance trust, overly human-like agents risk invoking an uncanny-valley effect, which can diminish user comfort and collaboration \cite{seeber_machines_2020}. Moreover, designing AI teammates that are both effective and acceptable to users remains a challenge, necessitating research into task-based and social factors that contribute to user acceptance \cite{oneill_21st_2023, stowers_improving_2021}. As AI tools evolve into partners in human-AI collaboration, it becomes essential to rethink the nature of this collaboration and the expectations surrounding it.

To engage with these research opportunities in the Human-AI collaboration space, our research focuses on the mental models of early adopters of multi-agent generative AI tools, emphasizing the developers' perspectives during the early stage of AI tool design. This focus is informed by the work of Bansal et al. \cite{bansal_beyond_2019} and Weisz et al. \cite{weisz_design_2024}, which highlight the crucial role that mental models play in shaping user interactions with AI systems. They argue that mental models influence how users interpret AI capabilities and limitations, which in turn affects their decision-making processes and collaboration outcomes within human-AI teams. By collecting data on the different conceptualization of multi-agent generative AI systems, we analyze these interactions from both AI-dominant and AI-assisted collaboration viewpoints \cite{yue_impact_2023}. \textcolor{black}{This paper builds on a design-oriented study of early adopters' workflows and tool development with multi-agent AI systems \cite{naik_designing_2025}, but the current study shifts the focus to how these early adopters conceptualize their collaboration with AI agents---specifically through the metaphors they use, the mental models they construct to navigate associated challenges, and their perceptions of transparency within these collaborative interactions. The primary research question our study is trying to answer is: \textit{How do early adopters of multi-agent generative AI systems conceptualize, interpret, and navigate collaboration with multiple agents?}}

\textcolor{black}{Our paper makes the following contributions to the CSCW and HCI communities:}
\begin{enumerate}
    \item \textcolor{black}{We present an empirical study of early adopters' mental models of multi-agent generative AI systems, focusing on how users conceptualize agent roles, manage collaboration, and navigate transparency and control in these complex environments.}
    \item \textcolor{black}{We extend existing human-AI interaction discourse by identifying challenges unique to multi-agent generative AI, such as distributed autonomy, inter-agent coordination, emergent behavior, and the need for layered transparency.}
    \item \textcolor{black}{We offer design insights specific to improving human-AI collaboration and teaming that highlight the limitations of current explainability and control approaches, introducing requirements like inter-agent traceability, collective reasoning visibility, and role-based customization.}
    \item \textcolor{black}{We reveal the metaphorical framings users apply to make sense of agent collectives, which reflect a shift toward viewing AI as dynamic, role-diverse collaborators.}
\end{enumerate}

Our research on multi-agent AI systems opens up exciting avenues for future exploration, particularly in applying CSCW lenses to understand agent-to-agent interactions, with or without human involvement. A critical aspect for future research is the examination of human agency and control within human-AI teams. This includes examining the specific points within human-AI interactions where human involvement is essential and identifying strategies to cultivate critical thinking among human team members to mitigate risks of over-reliance or excessive skepticism. This study advances CSCW literature by identifying new collaborative dynamics enabled by capabilities of multi-agent generative AI systems, transparency focusing on the human-to-agent and agent-to-agent interactions, and human mediation in the continuum between autonomy and human control in the collaborative work within workplace settings.

\section{Related Work}
This section reviews relevant literature on human-AI teams and collaboration, mental models in human-AI collaboration, transparency, \textcolor{black}{and contextualizing what we mean by ``early adopters,''} especially for advanced AI systems.

    \subsection{Human-AI Teams and Collaborative Work} 
    Recent CSCW research has focused on conceptualizing and characterizing human-AI teams, recognizing AI systems as active collaborators rather than mere tools. Seeber et al. \cite{seeber_machines_2020} emphasize the importance of mutual understanding, joint decision-making, and adaptive behavior in human-AI collaboration,  \textcolor{black}{arguing that these qualities are essential for AI systems to function as credible and responsive partners in collaborative settings.} Similarly, \textcolor{black}{studies such as} Liao et al. \cite{liao_questioning_2020} \textcolor{black}{have examined how the design of AI, particularly its anthropomorphic features, can influence collaboration quality. Their findings suggest that when AI systems exhibit characteristics that users associate with human teammates, such as responsiveness and communicative cues, users are more likely to engage with them as partners. This aligns with the broader perspective that the perceived agency of AI systems significantly shapes human collaborative behavior.}
    
    \textcolor{black}{Another key thread in the literature focuses on how collaboration outcomes are mediated by both task characteristics and user expertise.} Prior studies \cite{dellermann_future_2019, berretta_defining_2023} emphasize that the effectiveness of human-AI teams is influenced by the specific tasks being performed and the expertise of the users involved. \textcolor{black}{Expertise, in particular, shapes the mental models users form about how AI should behave and what role it should play, which in turn influences the effectiveness of the collaboration.} \textcolor{black}{Beyond these contextual factors, researchers have also explored the antecedents and outcomes associated with successful human-AI teaming. Trust in AI systems, clear communication strategies, and well-defined roles emerge as critical elements that shape team performance and satisfaction} \cite{zhang_investigating_2023, berretta_defining_2023, zhang_i_2024}. \textcolor{black}{These factors are not static; they evolve with users' experiences and perceptions of the AI's capabilities and reliability.} 
    
    \textcolor{black}{Lastly,} the literature also explores various collaboration modes and the roles that AI can play within teams. \textcolor{black}{For instance,} studies \textcolor{black}{by Berretta et al.} \cite{berretta_defining_2023} \textcolor{black}{and Bucinca et al.} \cite{bucinca_proxy_2020} \textcolor{black}{differentiate between AI as assistants, peers, or even supervisors within teams. These insights highlight the importance of designing systems that can flexibly support different collaboration modes and task configurations. Our work builds on this foundation by investigating how early adopters of multi-agent generative AI tools conceptualize and engage with these diverse roles in practice, particularly through the lens of their mental models and expectations of collaboration.}

    \subsection{Mental Models in Human-AI Collaboration}
    \textcolor{black}{Mental models, users' internal representations of how systems work, are central to effective human-AI collaboration and are defined differently across disciplines. In HCI, they are seen as cognitive constructs that guide user interaction, with research emphasizing the importance of accessible system metaphors to support understanding \cite{nielsen_hiding_2007, villareale_understanding_2021}. In CSCW, the concept of mental models is often framed through the lens of shared cognition within teams, crucial for coordination and trust in human-agent systems \cite{phillips_tools_2011, schelble_lets_2022}. From the AI perspective, particularly in the area of explainable (XAI), mental models are linked to users' interpretations of system behavior, functionality, and transparency. Research has shown that cognitive biases, such as anchoring effects, can distort these interpretations and lead to flawed understanding of how AI operates \cite{nourani_anchoring_2021, kuang_collaboration_2023}. Grounded in design-oriented definitions of mental models (see, for example, \cite{norman_design_2013, kulesza_tell_2012, kulesza_too_2013}), this study adopts a conceptualization of mental models as dynamic, evolving constructs that shape user trust, system usability, and perceived predictability. Adopting the human-centered AI system design perspective, we focus on three key dimensions of mental models \cite{cerejo_understanding_2023}: (1) how to use the AI tool, (2) when to use it, and (3) what it can do. These dimensions are particularly salient when examining how early adopters make sense of multi-agent generative AI systems, where capabilities are both novel and often non-intuitive.}
    
    \textcolor{black}{Understanding the mental models in this context is important.} Past research shows that when users possess a well-formed mental model of an AI's capabilities and limitations, they are better equipped to make informed decisions regarding when to rely on AI recommendations and when to exercise their judgment \cite{bansal_beyond_2019}. This understanding is particularly vital in high-stakes environments, such as healthcare and finance, where the consequences of decisions can be profound \cite{bansal_beyond_2019}. Furthermore, studies have shown that discrepancies between human and AI mental models can lead to misunderstandings and suboptimal collaboration outcomes \cite{schemmer_meta-analysis_2022}. \textcolor{black}{Prior work has highlighted using the 3-Gap Framework \cite{kayande_how_2009} that effective collaboration relies on minimizing the gap between the AI model, the user's mental model, and the real-world context. Misalignment, especially between the system's logic and the user's understanding, can result in confusion, mistrust, and breakdowns in collaborative performance.} The role of mental models in human-AI collaboration is further complicated by the varying types of collaboration \textcolor{black}{configurations}, such as AI-dominant versus AI-assisted interactions. Research indicates that the type of collaboration influences how users attribute responsibility and trust to AI systems, which in turn affects their understanding about these AI systems \cite{yue_impact_2023}. For instance, in AI-dominant scenarios, users may develop a more passive mental model, relying heavily on AI decisions \cite{passi_appropriate_2024}, while in AI-assisted contexts, users may maintain a more active role, leading to a more nuanced understanding of the AI's capabilities \cite{yue_impact_2023}.
    
    \textcolor{black}{These insights from literature showcase the importance of investigating how mental models are formed when using complex AI systems. In our study of emerging ``multi-agent generative AI system,'' where the boundaries of human and AI roles are increasingly fluid, this understanding can help define those roles more clearly and contribute to the discourse of human-AI collaboration in complex systems.}

    \subsection{Transparency in human-AI collaboration}
    \textcolor{black}{While agent transparency has become a bridge to better mental models by supporting human understanding of AI behavior, most of this work has centered on single-agent AI systems \cite{vossing_designing_2022}. In contrast, multi-agent generative AI systems, in which multiple autonomous agents interact, often in emergent and unpredictable ways, is still a growing area of research. Despite their growing adoption, especially among early users of these collaborative generative AI tools, little is known about how humans interpret, engage with, or build trust in these complex AI ecosystems.} 

    \textcolor{black}{Transparency} through explainability is a key factor in fostering user trust, which is crucial for effective collaboration between humans and AI \cite{oneill_21st_2023}. \textcolor{black}{In single-agent contexts,} explanations \textcolor{black}{that are} accessible and tailored to diverse user types, focusing on their \textcolor{black}{varying levels of expertise, have been shown to improve transparency and user comprehension} \cite{ribera_can_2019, kuhl_you_2019}. \textcolor{black}{These insights have shaped the development of} human-centered XAI, \textcolor{black}{which emphasizes clarity, interpretability, and adaptability of system design} \cite{liao_human-centered_2022}. \textcolor{black}{However, it remains unclear how these principles transfer to multi-agent generative AI systems, where the sources of decisions and outputs may be distributed across multiple interacting agents.}
    
    \textcolor{black}{A recent empirical study by Vössing et al. \cite{vossing_designing_2022} examined how agent transparency, providing insight into an AI system's reasoning and uncertainty, affects trust and task outcomes in an AI tool. This study highlights the importance of calibrated and context-aware explanation design, though its insights are grounded in a single-agent domain-specific setting. Its applicability to multi-agent generative systems, where agency is distributed and explanations must account for inter-agent dynamics, remains an open question.} Such approaches help correct user misconceptions and adapt mental models, fostering better understanding and trustworthiness \cite{gero_mental_2021}. In the context of human-AI interaction, trust is challenging in AI systems, given the complexity and emergent nature of agent interactions \cite{benk_two_2023}. Effective trust-building strategies include incorporating human-in-the-loop designs, where users can actively guide AI behavior and provide feedback, thus fostering a sense of control and adaptability \cite{kaur_trustworthy_2022}. Managing user expectations is crucial for trust development. Zhang et al.\cite{zhang_i_2024} found that the quality and type of explanations influence user trust and reliance on AI systems. Another key thread in the literature focuses on how the design of AI systems influences users' mental models, their internal representations of how the system works. XAI has been shown to foster more accurate and usable mental models by making decision-making processes more transparent \cite{schemmer_meta-analysis_2022, berretta_defining_2023, zhang_i_2024}. In contrast, opaque systems can lead to confusion, eroded trust, and poor collaboration outcomes.
    \textcolor{black}{These constructs around transparency, trust, and explainable AI provide a foundation for human-AI collaboration, and are largely rooted in single-agent paradigms. Through our study, we contribute to these transparency considerations by beginning to extend them to multi-agent generative AI systems.}

    \textcolor{black}{Given that these insights stem from studies on single-agent generative systems or multi-agent systems without generative capabilities, there remains a pressing need to understand how these challenges manifest in the context of multi-agent generative AI systems, a rapidly emerging class of tools not yet widely adopted but growing in use. As these systems introduce new forms of interaction, coordination, and agency among multiple AI entities, it becomes essential to examine how early adopters engage with them. In our study, we focus on developers as both early adopters and active users of these systems, whose mental models offer valuable insights into potential pitfalls and challenges faced when interacting with multi-agent generative AI tools.}

    \subsection{\textcolor{black}{Contextualizing early adopters: developers shaping multi-agent generative AI}}
    \textcolor{black}{Developers often serve as early adopters of emerging AI technologies, providing essential insights into their usability, integration potential, and scalability. Here, the term ``developer'' includes individuals who can be of any roles (such as technical and non-technical) engaging in development-related activities, thereby shaping the trajectory of technological adoption and refinement. In this capacity, they function as \textit{de facto} designers, influencing the evolution of these technologies through their interactions and feedback \cite{kumar_key_2020}. Their knowledge of AI systems---including their purpose, functionality, and broader implications---is critical for identifying barriers to adoption and ensuring transparency in multi-agent generative AI systems. Prior research on developers' engagement with programming tools and intelligent systems, such as self-tracking technologies \cite{choe_understanding_2014}, demonstrates how they develop tailored solutions to address system limitations, effectively bridging design gaps. For example, studies by Kulesza et al. \cite{kulesza_tell_2012} highlight how developers' mental models and understanding of AI systems contribute to the formulation of more effective explanation strategies, ultimately facilitating adoption and reducing cognitive load. Early adopters of AI technologies often play a pivotal role in shaping integration of these systems within organizations. Their experiences can provide valuable insights into the development of mental models that facilitate effective human-AI collaboration. For instance, research has demonstrated that early adopters tend to have a more nuanced understanding of AI capabilities, which allows them to leverage AI tools more effectively in their workflows \cite{beghetto_new_2023}. This is supported by findings that suggest early adopters are more likely to engage in reflective practices that enhance their mental models, thereby improving their collaborative outcomes with AI systems \cite{beghetto_new_2023}. Given their hybrid role as users and system shapers, early adopters provide a valuable lens through which to examine the evolving dynamics of human-AI collaboration, especially in multi-agent generative AI settings where workflows and agent roles are still being defined. Their experiences and practices can therefore better inform the design and development of these systems, ensuring they are both effective and accessible to a wider audience.}

    \textcolor{black}{This perspective motivates our focus on early adopters of multi-agent generative AI systems in this study. In the following sections, we detail the methods used to investigate their mental models around conceptualizations of these human-AI teams, collaborative strategies and the challenges faced, perceptions of transparency, and design preferences or patterns with AI agents in dynamic, real-world contexts.}

\section{Methodology}
\textcolor{black}{Given the emerging nature of multi-agent generative AI systems, we adopted an interpretivist approach \cite{creswell_qualitative_2018}, conducting semi-structured interviews centered on a few reflective tasks, such as a tool walkthrough, followed by the use of cognitive maps and thematic analysis of the data to explore how key constructs--- mental models, collaboration strategies, transparency--- manifest in real-world usage. Rather than applying a single, predefined framework, we drew from multiple areas of literature to inform our sensitizing concepts while we generated themes inductively. Our findings offer an integrative perspective that connects these domains, which we elaborate in the discussion as a basis for extending knowledge of human-AI collaboration.} To elicit the mental models of early adopters, we interviewed \textcolor{black}{13} participants using a semi-structured interview protocol that included retrospective tasks and reflection to prompt participants to share previous experiences with the tools and processes they use to develop or design multi-agent generative AI technologies. The interview topics covered how the participants understand, interpret, and explain the multi-agent generative AI they develop and interact with, reasons of using these tools, their approaches to transparency in such system development, and any challenges they encountered. The retrospective tasks and reflection activities included an in-depth exploration of their tool's use-cases using different multi-agent AI frameworks, as well as their understanding of the system.
    
\subsection{\textcolor{black}{Participant Recruitment}} We tried to recruit the participants for the interview through 3 mechanisms: 
(1) a follow-up screener survey sent to the 18 participants of a prior pilot survey, which had originally been distributed to online communities of generative AI developers who communicate on Reddit (r/OpenAI, r/AutogenAI, r/ArtificialIntelligence, r/agi) and Discord (AutoGen); (2) recruitment emails distributed across Microsoft; and (3) snowball sampling techniques with both groups. \textcolor{black}{Of the 18 individuals contacted through the follow-up screener survey, which targeted prior pilot survey participants in the external communities, only one response was received. In contrast, recruitment emails sent within Microsoft, along with snowball sampling from the company group, yielded 13 participants. Due to the limited response from external communities of early adopters (n = 1), we refined the scope of the study subsequently to focus exclusively on workplace settings within the Microsoft. As a result, we excluded the one response received from the external communities, bringing the final participant count to 13.} The primary inclusion criteria for participation required individuals have first-hand experience creating or experimenting with a multi-agent AI framework within their professional roles and possess an intermediate or greater level of knowledge about multi-agent generative AI. Inclusion criteria for this study required participants to be English-speaking and at least 18 years of age or older. Participants \textcolor{black}{from the internal distribution lists} were excluded if they had higher executive roles in industry, to avoid any potential bias from senior decision-makers.

\subsection{\textcolor{black}{Participant Details}}
\textcolor{black}{A total of 13 participants who were all employed at Microsoft individually participated in the study (Table ~\ref{tab:demographics}). Prior to participation, all individuals were provided with an informed consent form, and only those who signed the form were included in the study. Participation was entirely voluntary, and no compensation was offered. At the time of the study, all researchers except one were affiliated with Microsoft. As researchers affiliated with the same organization as the participants, we had no prior professional or personal connections with the participants and had not interacted with them before the commencement of the study.} \textcolor{black}{Five} participants were involved in non-technical job roles, \textcolor{black}{such as product managers, designers etc.,} and eight were associated with technical job roles, \textcolor{black}{such as software developers, data scientist etc. In terms of overall professional experience, the participant group (N = 13) represented a diverse range. Four participants had 3–5 years of experience, six had 6–10 years, one had less than 1 year, one had 11–15 years, and one had 16–20 years of professional experience.} The participants were primarily based in the United States and Europe. \textcolor{black}{Participants were not only involved in developing proof-of-concept (POC) and production-level multi-agent generative AI systems, but were also active users of these tools, primarily for individual or team use in their day-to-day tasks.} These tools were built using various frameworks, including but not limited to AutoGen \cite{wu_autogen_2023, dibia_multi-agent_2023}, Semantic Kernel, TaskWeaver \cite{qiao_taskweaver_2023}, BizChat, Sydney, LlamaIndex, Oagent, and several custom-built frameworks. They reported that the primary motivations for developing these tools were to simplify their work, learn about the technology, understand its capabilities and limitations, experiment, and share knowledge with others.

\begin{table*}
\textcolor{black}{
  \caption{\textcolor{black}{Participant's demographics in the order of their participation}}
  \label{tab:demographics}
  \begin{tabular}{p{0.10\linewidth} p{0.15\linewidth} p{0.20\linewidth} p{0.15\linewidth}}
    \toprule
    \textbf{Participant} & \textbf{Role} & \textbf{Experience Level} & \textbf{Usage Type} \\
    \midrule
    P1 & Technical & 3-5 years & Individual use \\
    P2 & Technical & 6-10 years & Individual use\\
    P3 & Non-technical & 3-5 years & Team use \\
    P4 & Technical & 6-10 years & Team use \\
    P5 & Technical & 11-15 years & Individual use\\
    P6 & Technical & 16-20 years & Team use \\
    P7 & Non-technical & Less than 1 year & Individual use \\
    P8 & Non-technical & 6-10 years & Team use\\
    P9 & Technical & 6-10 years & Team use\\
    P10 & Technical & 6-10 years & Individual use \\
    P11 & Non-technical & 3-5 years & Team use\\
    P12 & Technical & 3-5 years & Team use \\
    P13 & Non-technical & 6-10 years & Individual use\\
    \bottomrule
  \end{tabular}
}
\end{table*}

\subsection{\textcolor{black}{Procedure}}
\textcolor{black}{Participants took part in individual, remote interview sessions conducted via Microsoft Teams during their regular working hours. Each session lasted between 45 to 60 minutes and was conducted by the first author. The interviews began with a brief set of warm-up questions focused on participants' day-to-day work and interests, followed by the main set of semi-structured interview questions. 
We designed the interviews to elicit participants' reflections on tasks involving multi-agent AI frameworks and their conceptual understanding of such systems based on the three key dimensions of mental models \cite{cerejo_understanding_2023}---how to use, when to use, what it can do. The interview protocol was structured to explore key themes, including participants' motivations for adopting multi-agent AI systems, their transparency practices, and the challenges encountered during development and deployment. We kept the questions open-ended to encourage detailed, reflective responses using tool walkthrough. Specific areas of inquiry included: }
    
\begin{enumerate}
    \item \textcolor{black}{\textbf{Conceptual Understanding:} The participants were asked, \textit{``Based on your knowledge, can you describe multi-agent AI?''} to assess their understanding and mental models of these systems.}
    \item \textcolor{black}{\textbf{Practical Applications:} To surface concrete use cases on how they use these tools, participants were invited to describe scenarios of use in their daily work, with prompts like, \textit{``Without revealing any business-confidential projects, can you walk me through how you use this tool in day-to-day tasks?''}}
    \item \textcolor{black}{\textbf{Capabilities and Limitations:} Participants reflected on their experiences using these systems through questions such as, \textit{``Based on your experience, what are some capabilities and limitations of using this tool?''} to understand when they use and what these tools can do.}
    \item \textcolor{black}{\textbf{Transparency Practices:} To explore participants' conceptualizations of transparency, they were prompted with questions like, \textit{``What comes to your mind when I say `transparency practices in multi-agent AI'?''}}
\end{enumerate}

\textcolor{black}{The semi-structured format allowed for flexibility in probing participant responses, facilitating a rich exploration of diverse perspectives and real-world applications of multi-agent generative AI systems. We video recorded all the sessions for later analysis. This study was reviewed and approved by the Institutional Review Board (IRB) of the Microsoft.}

\subsection{\textcolor{black}{Analysis}}
\begin{figure}
    \centering
    \includegraphics[width=1\linewidth]{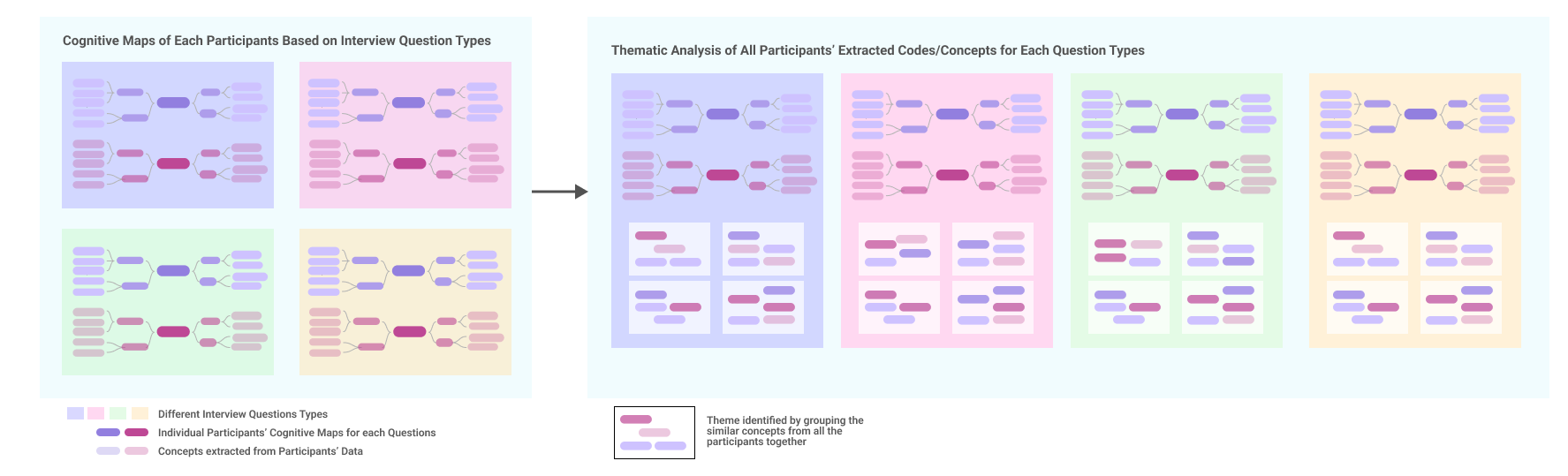}
    \caption{\textcolor{black}{The visual diagram of our analysis process. We first conducted cognitive mapping separately for each participant across each question type (first half), then used concepts identified from these maps to conduct thematic analysis (second half)}}
    \label{fig:analysis}
\end{figure}

\textcolor{black}{Approximately 9.0 hours of recorded data were collected through interview sessions with early adopters. Two researchers transcribed the recordings using the organization's internal transcription tool and manually editing wherever needed. The transcribed interviews were then cleaned and consolidated into a single document, structured according to the interview questions. Responses to warm-up questions regarding participants' background information were excluded from the subsequent analysis.
To examine participants' mental models from the human-centered design perspective, we generated cognitive maps for each individual based on their responses to the interview questions (Figure ~\ref{fig:analysis}). Cognitive mapping, a technique used to represent individuals' or groups' mental models visually, facilitates the analysis of complex qualitative data \cite{carley_extracting_1992}.
Following the creation of the cognitive maps, we (team of two researchers) then conducted thematic analysis \cite{braun_thematic_2012} to identify recurring patterns, divergences, and conceptual tensions across participants' responses for each interview question from the 89 extracted codes/concepts in the cognitive maps. Through this inductive analysis, we generated key themes that highlighted how early adopters conceptualize multi-agent AI systems, characterize such systems according to collaboration types, articulate transparency practices, define the roles of AI agents, and express their design preferences. For instance, in response to the question concerning conceptualizations of multi-agent AI systems, 16 codes from seven participants were synthesized into four dominant themes: ``a team of specialized agents,'' ``divide and conquer model,'' ``human-like agents,'' and ``software component analogies.''
Over the course of seven weeks, three researchers held frequent meetings and engaged in group discussions to evaluate the validity and reliability of the themes identified for each interview question, and iteratively reorganized the analysis space.}

\section{Findings}
In this section we present the themes we found across our interviews with early adopters (developers) of multi-agent Generative AI systems. We start by outlining some metaphors or analogies our participants used for conceptualizing the multi-agent generative AI systems in this human-AI collaboration. \textcolor{black}{Next, we explain how our participants talked about using these AI tools, focusing on what the tools could and couldn't do, depending on the type of collaboration they were involved in} \cite{yue_impact_2023}. We then describe how our participants discussed the perception of transparency when designing and developing such systems. 

\subsection{Mental Model Analogies and Frameworks for Conceptualizing Multi-Agent Generative AI Systems}

\begin{figure}
    \centering
    \includegraphics[width=1\linewidth]{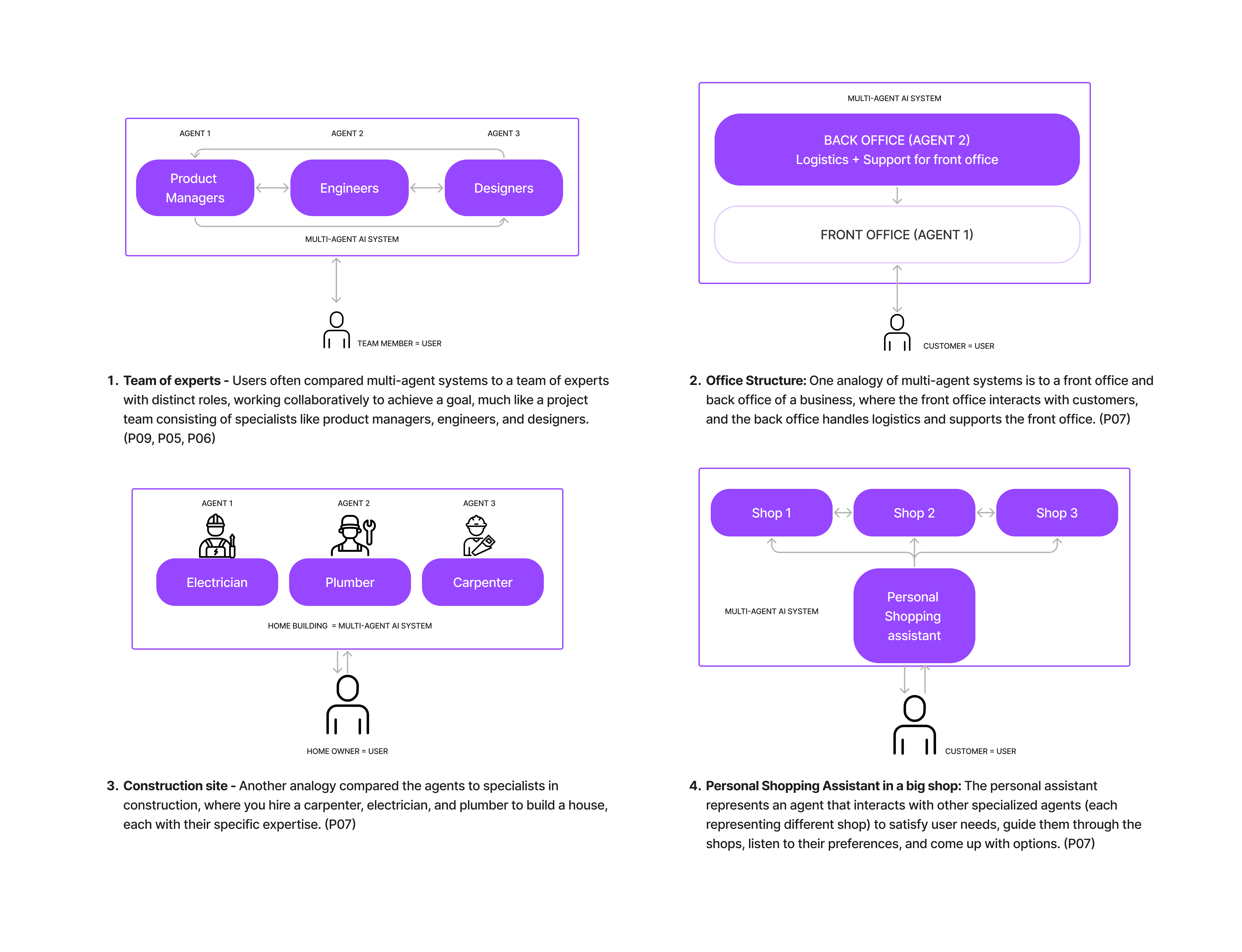}
    \caption{Different mental model analogies of conceptualization of using multi-agent generative AI tools}
    \label{fig:metaphors}
\end{figure}

The mental models of multi-agent Generative AI held by the early adopters reflect diverse conceptualizations ranging from simple analogies to sophisticated descriptions of system dynamics. These descriptions include both technical and social analogies, often overlapping. The metaphors used to describe this human-AI collaboration are (as shown in Figure ~\ref{fig:metaphors}): (1) A ``team of experts'' scenario, wherein AI agents assume distinct roles, such as project manager, engineer, and designer, collaboratively working towards a common objective; (2) A ``front and back office'' model, where humans engage with the AI in a front-office capacity, while the agents further interact with a back office to fulfill user requirements; (3) A ``construction site'' analogy, likening collaboration to building a home where each AI agent fulfills a specialized role (e.g., plumber, electrician, carpenter); and (4) A ``personal shopping assistant'' framework, in which an AI agent functions as a personal assistant, addressing the needs of a user seeking specific products and coordinating with other agents to meet these needs.

In this collaboration, the most common patterns or features we found include: (1) thinking in terms of a team comprised of multi-disciplinary specialists; (2) thinking in terms of how a task is divided and conquered; (3) anthropomorphizing agents to have human characteristics; and (4) comparing multiple agents to how components within software function together.

\subsubsection{Collaboration within a team of specialized agents}
Many participants conceptualized multi-agent AI as a collection of independent but collaborating AI agents, each performing a distinct role. This model mirrors human teams where members contribute specific skills.
Similarly, P09 compared agents to team members with different skills: {\itshape ``We might have a product manager, we might have an engineer, and everybody has their own experiences, their own skills, their own tools, that they bring to the table to work together.''} In these approaches, the collaborating agents have specific roles---which the other agents are able to know about---but the specific mechanisms of collaboration are not necessarily relevant. Users created {\itshape role-based agents} with distinct roles, such as travel agents, event managers, analysts, managers, data scientists, and assistants, and {\itshape task-based agents} according to the sub-tasks assigned like summarizers, reviewers, writers, and phishing detectors. These agents conversed among each other to automate content creation like release notes and newsletters (P01, P02), streamline processes within platforms like GitHub and Teams (P04, P10), analyze potential phishing emails (P05, P13), and automate nurse checkups (P12). These multi-agent systems enabled the distribution of tasks among specialized agents, mimicking collaborative human interactions and demonstrating significant potential to improve efficiency, creativity, and decision-making across different workflows.

\subsubsection{Divide and Conquer Model}
The divide and conquer metaphor was reflected in how participants described multi-agent AI approaches in terms of how they enhance efficiency by breaking down complex tasks into smaller, manageable sub-tasks that are each handled by an agent specifically suited for that task. 
Participant P05 emphasized that, {\itshape ``[...] if you can break the task to a smaller one, there's a higher chance it can be completed.''} P07 explained: {\itshape ``The main point of multi agent is that compared to having submitting a prompt to one agent and expecting it to be able to perform any complex task, you're now partitioning it, where each agent is a specific, intelligent component as a very well defined and confined set of skill set, tools and abilities that can perform and together all these agents, they orchestrate and coordinate with each other to accomplish a complex task.''} This metaphor emphasizes the collaboration mechanism (identifying sub-tasks and then delegating them to agents that can perform those sub-tasks), but the specialized-ness of the agents is flexible. In other words, agents can be ``specialized'' in terms of how they accomplish that specific task or in terms of how they approach those types of tasks, in general. 

\subsubsection{Human-like Agents}
Within this conceptual framework, agents are perceived as emulating human characteristics, such as possessing identities, having a capacity for memory that is described in human terms, engaging in planning activities, and being able to interact naturally. The collaboration possibilities for how these agents with different identities interact are described in diverse ways. Participant P07 uses a customer satisfaction analogy: {\itshape ``Configuration is like you have front and back office situation and a customer to satisfy and each office has their own responsibility, but work in a different way.''} Here, the ``front office'' describes the AI agent(s) the ``customer'' interacts with directly, while the ``back office'' agent(s) interact and perform their responsibilities out of sight of the customer, but all of these agents posess an implied desire to satisfy the customer.

Similarly, Participant P13 personifies agents saying: {\itshape ``You are giving each one of these LLM agents personality, a role, and a specific task or duties to perform. You're basically creating a team of little experts that can do certain things.''} Each of these experts have different abilities like humans and they converse with each other by passing context back and forth to achieve a shared goal.

\subsubsection{Software Components}
Some participants described these multi-agent systems using language found in existing software paradigms, such as microservices or modular functions, suggesting that the concept is not entirely new but an extension of familiar patterns. Participant P04 explained, {\itshape ``[...] it's just like software. The more you can create very small services, really small functions that have a single input and a very predictable output.''} Participant P03 suggests reusing different agents similar to software components: {\itshape ``Build your agents once and then dynamically reuse them to achieve greater tasks.''} With this approach, delegating a task to an agent, whether the task is delegated by a human or by another agent, is conceptually similar to calling a function with the instructions for that task as an input. However, as we describe in the next section, the non-deterministic nature of working with multiple AI agents is likely to result in non-``predictable output[s].''

\textcolor{black}{These diverse metaphors and conceptual patterns illustrate how early adopters actively construct mental models to make sense of their collaborations with multi-agent generative AI systems. Drawing from the human-centered design definition of mental model, we understand these mental models not as static representation but as dynamic, experience-based interpretations that guide users' expectations, decision-making, and interaction with technology. This framing is relevant in the context of an emerging and rapidly evolving technology like multi-agent generative AI, where formal models or established collaboration paradigms are limited. By centering mental models in our analysis, we were able to surface how users negotiate complexity, ambiguity, and the division of roles in human-AI collaboration. Thus, this mental model lens enabled us to highlight how meaning-making unfolds in real-world use, offering insights not only into user experience but also into implications for system design.}
            
\subsection{Multi-Agent Generative AI System Characteristics and User Collaboration Type}
In this section we expand on the tensions our participants discussed that exist between collaborations that are AI-dominant and collaborations where there is more human involvement. 

\subsubsection{AI-dominant collaboration} 
AI dominant collaboration models prioritize agent autonomy, often allowing AI agents to operate independently with minimal human intervention \cite{yue_impact_2023}. Through our analysis, we identified several AI-dominant collaboration patterns from our participants' descriptions of multi-agent generative AI tools, centering on concepts such as orchestration, autonomy, and control.

Conversable agents with specialized roles, autonomously orchestrated to enhance efficiency, makes the system highly AI-dominant. For instance, P06 described a setup where agents focused on particular tasks, such as data analysis or decision-making, independently managing their responsibilities within an overarching task framework. These agents converse among themselves and make decisions independently. This role specialization within autonomous systems can support complex workflows, minimizing direct human intervention to achieve targeted goals. P11 expressed concerns over the stochastic and unpredictable nature of such an approach, arguing that it would affect reliability, especially for high-stake applications. 

P04 and P13 expressed an acceptance of how relying on AI agents that generate non-deterministic outputs can shift such collaborations into a more AI-dominant mode. By allowing agents a high degree of autonomy, this approach enables flexibility and creativity. However, the complexity of interactions among agents can lead to unpredictability, a phenomenon described by P07 as contributing to the system's non-deterministic, sometimes spontaneous, behavior. Technical users like P11, however, voiced a preference for determinism, suggesting that AI-dominant models may benefit from structured guidance to balance creativity with control.

Extending beyond conventional \textcolor{black}{single agent} AI functions like question-answering, enabling fully automated workflows can also shift such collaborations into AI-dominant modes. P13 described an example of multi-agent systems autonomously generating content across various media platforms, integrating tasks such as transcription, summarization, and image creation. Such workflow automation requires minimal human input, representing a streamlined, high-autonomy system design.

\subsubsection{AI-assisted collaboration} AI-assisted collaboration emphasizes a partnership where human oversight and control are essential components in achieving predictable, structured outcomes \cite{yue_impact_2023}. Participants favoring this approach, like P11, highlighted a preference for direct human orchestration over autonomous AI control, enabling clearer task management and reducing randomness in complex workflows. Technical users (e.g., P02, P05, P11) also emphasized the importance of deterministic outputs, where clear prompt structuring and orchestration minimizes cognitive load by maintaining a stable framework and ensures reliability. In these systems, developers act as human orchestrators, much like how P02 described breaking down tasks up above, who closely manage task decomposition, which supports controller, user-directed workflows that optimize accuracy and efficiency.

\subsubsection{Challenges in both collaboration types} Our participants described several underlying challenges that are prevalent in both AI-dominant and AI-assisted collaborations, including error propagation, hallucinations, unproductive agent loops, unpredictability, opaque processes, and non-standard practices.
 
\textbf{Hallucinations and context management issues} present persistent obstacles, affecting both AI-dominant and AI-assisted models. P06 cited cases where agents generated inaccuracies or lost contextual relevance, leading to compounded errors, especially when one agent's error feeds into another.Such cascading errors are especially problematic in AI-dominant configurations, where multiple autonomous agents operate with minimal human oversight. This inherent characteristics of advanced AI systems can erode the user trust, thus emphasizing the importance of mechanisms that foster traceability or error management.

Participants (e.g., P06, P13) observed that agents sometimes became ``stuck'' in repetitive, unproductive cycles, which offered little to no added value to the collaborative task at hand. These ``agent loops'' were particularly pronounced in AI-dominant environments, where agents might autonomously continue a cycle without human intervention until the inefficiency became noticeable. Participants (e.g., P03, P09) observed that the unpredictability of agent behavior compounded these challenges; as the number of agents increased, so did the complexity of possible pathways, leading to exponential growth in workflow management challenges. This \textbf{unpredictability and unproductivity} underscores the need for human intervention at critical decision points as ``user-defined checkpoints'' or ``structured control points'' to prevent compounding inefficiencies and unintended consequences within the autonomous processes.

The \textbf{lack of transparency} in agent behaviors and decision-making processes emerged as a significant barrier in both AI-dominant and AI-assisted settings. Participants (P06) described many agents as operating in a ``black-box'' manner, where the opacity of underlying processes complicated efforts to trace or understand agent actions. This lack of interpretability was further exacerbated by the \textbf{absence of standardized practices}, which made it difficult to integrate agents cohesively within established workflows. Participants (P04) emphasized the need for visual and interpretative tools that could provide developers with deeper insights into agent behaviors, aiding debugging and enhancing performance evaluation. Such tools are essential for providing developers with traceability and control in multi-agent systems, particularly when autonomous agents operate with significant independence.

\textcolor{black}{The challenges and collaboration patterns described across AI-dominant and AI-assisted collaboration types highlights the importance of users' underlying mental models in shaping how they engage with multi-agent generative AI systems. Drawing on the human-centered design definition, we conceptualized mental models around collaboration types as users' evolving experience-based understandings of how these system function --- what agents are capable of, how they relate to one another, and where human oversight is needed. These models guide expectations, influence orchestration strategies, and inform how users manage risks like unpredictability or error propagation.}

\subsection{Perceptions of Transparency}
Through the thematic analysis of the data we were able to identify some common patterns among participants' perceptions of transparency in multi-agent AI systems. These patterns highlight the consensus on the importance of transparency for an effective human-AI collaboration, with factors like trust, verifiability, visibility, explainability, and the ability to understand potential data leaks permeating our participants' descriptions.

Participants like P06 explicitly mentioned the role of transparency in building user \textbf{trust}. They noted, {\itshape ``Transparency is paramount to creating trust. If users can see what the system's doing and understand how it operates, they are more likely to trust it.''} This emphasis on transparency as foundational to trust-building was echoed by P05, who also highlighted the need to clearly explain {\itshape ``how the system works, the pros and cons, and the risks involved.''}

Several participants, particularly those with technical backgrounds, highlighted transparency as crucial for debugging and system improvement. For instance, P03 emphasized that developers need to see {\itshape ``the inner workings''} of the agents to be able to \textbf{verify} answers or trace errors. P04 elaborated on the need for a transparent user interface that shows how agents interact, suggesting a group chat-style interface or a directed graph to depict the flow of information in a \textbf{clear visualization}. P11 also emphasized the importance of tracking what information agents retrieved and the reasoning behind their decisions, to ensure an event-driven orchestration can be understood, verified, and/or explained by a developer attempting to maintain oversight for any collaboration type.

Transparency was also seen as necessary for \textbf{preventing misuse, errors, and data leaks}, especially in complex multi-agent environments. P07 provided an example where agents interacted with confidential documents, stating that transparency {\itshape ``could allow you to detect potential data leaks or biases,''} especially in a multi-agent setup where information flows through different stages. Similarly, P13 emphasized the importance of incorporating critical evaluation metrics for AI-generated content to determine if any bias or corruption was present in the agents.

\textcolor{black}{These insights underscore the centrality of transparency in shaping users' mental models of multi-agent generative AI systems. Participants' emphasis on trust, verifiability, and oversight reflects how transparency functions as a mechanism for understanding and managing system behavior. In complex, distributed collaborations, transparency becomes critical for users to maintain control and confidence. By highlighting this, our study extends CSCW work on human-AI collaboration to account for the unique challenges of multi-agent settings, where transparency directly influences how users come up with strategies to better coordinate, debug, prevent error/misuse, and evaluate AI-driven workflows.}

\section{Discussion}

\textcolor{black}{This study examines how early adopters engage with multi-agent generative AI tools, with a focus on their mental models around collaboration dynamics, transparency, and control strategies that emerge in these uniquely complex systems. As generative AI moves beyond single-agent interactions, understanding how users conceptualize and coordinate with multiple, semi-autonomous agents becomes increasingly critical to both theory and design. Building on human-centered design approaches to mental models \cite{cerejo_understanding_2023}, we explored not just what users think these systems do, but also how and when they believe they should be used by highlighting the patterns of use and AI roles, challenges in collaboration, different perception of transparency, and their design preferences. These dimensions are central to ensure meaningful and effective human-AI collaboration.}

\textcolor{black}{Participants developed distinct \textit{metaphorical framings} to make sense of multi-agent systems, describing them as ``teams of experts,'' ``construction sites,'' or ``personal assistants.'' These metaphors suggest a shift from viewing AI as monolithic tools to seeing them as \textit{role-diverse collaborators}. While our findings asserted existing collaboration paradigms -- AI-dominant and AI-assisted configurations \cite{yue_impact_2023} -- we found that multi-agent contexts extend these categories in significant ways. In AI-dominant configurations, user control is minimal, and agents operate autonomously \cite{yue_impact_2023}. However, in multi-agent systems, this autonomy becomes distributed across specialized agents, introducing new coordination challenges and risks, such as compounding errors and inter-agent misalignment. In contrast, AI-assisted configurations, typically associated with human oversight, become more nuanced, with users engaging in selective, layered supervision of individual agents or agent clusters. This more targeted control enables task-specific debugging and oversight, but also increases cognitive and interface demands. Across both the configurations, participants also emphasized the need for \textit{multi-layered transparency} and \textit{distributed autonomy}, not only to interpret individual agents' behaviors but also to trace how agents interact and influence one another. This layered visibility was seen as critical for debugging, trust-building, and managing the unique risks posed by \textit{inter-agent coordination} and \textit{emergent behavior} like error propagation. By foregrounding mental models in our analysis, we contribute to CSCW literature by extending the human-AI collaboration discourse beyond dyadic (human-single AI) settings, offering a lens into how users navigate coordination, autonomy, and control in more complex multi-agent generative environments. This perspective offers a step toward developing more standardized practices for engaging with such systems.}

\textcolor{black}{Different user groups approached multi-agent collaboration with varied expectations. Technical users favored detailed control and transparency features, including the ability to track agent-to-agent exchanges, while non-technical users preferred more structured guided workflows, simplified oversight, and predictable outcomes. Despite these differences, all participants expressed a need for better mechanisms to manage and monitor agent coordination, indicating that the complexity introduced by multiple generative agents requires new forms of oversight and design scaffolding beyond those common in single-agent AI systems. These insights emphasize understanding users' situated goals, expectations, and heuristics as part of a broader human-centered mental model framework \cite{cerejo_understanding_2023}.}

\textcolor{black}{Studying early adopters of multi-agent generative AI allowed us to capture evolving user expectations, conceptual models, and workarounds at a moment when the human-centered design space for advanced AI systems remains under-defined. Our qualitative, exploratory approach was effective to unpack the situated, context-dependent ways users engage with these novel systems, offering first-hand insights into how multi-agent configurations are experienced, interpreted, and operationalized in practice.}

\textcolor{black}{While prior research on mental models, explainability, and human-AI teaming has provided key foundations \cite{bansal_beyond_2019, weisz_design_2024, ferrario_ai_2020}, these studies often assume single-agent contexts. Our findings extend this literature by surfacing dynamics that are specific to multi-agent generative AI, such as distributed autonomy, inter-agent dependency, and the need for configurable role structures. These insights build on and refine existing theories of human-AI collaboration \cite{liao_human-centered_2022, seeber_machines_2020}, suggesting a need to evolve current frameworks to better account for the distributed, generative nature of agent collectives.}

\textcolor{black}{While the importance of transparency and explainability in AI systems is well-established in past research works \cite{vossing_designing_2022}, our findings suggest that multi-agent generative AI introduces distinct challenges that go beyond those seen in single-agent. Specifically, early adopters emphasized the need to understand not just \textit{what} individual agents are doing, but \textit{how} agents interact, influence each other's outputs, and contribute to system-wide behavior. This requires new forms of \textit{inter-agent traceability} and \textit{collective reasoning visibility}, which are not typically addressed by conventional explainability approaches. Similarly, issues such as role-based customization, oversight at different layers of control, and coordination across agents take on added complexity in multi-agent systems, where autonomy is distributed rather than centralized. Systems must support such customization, traceability, and scalable transparency, while balancing usability for both technical and non-technical users. Without these adaptations, multi-agent tools risk overwhelming users, reducing trust eroding their confidence \cite{oneill_humanautonomy_2022}, limiting adoption, and promote unintended behavior. These findings highlight the importance of new collaboration scaffolds and reinforce the value of grounding system design in a human-centered understanding of users' evolving mental models \cite{cerejo_understanding_2023}.}

\textcolor{black}{By focusing specifically on multi-agent generative AI systems, this study reveals how early adopters are not merely using these tools but actively shaping new collaborative paradigms. Their interaction patterns, metaphors, and expectations surface a pressing need to rethink how we design, explain, and govern complex generative systems composed of many agents. In doing so, we extend existing models of human-AI interaction to account for the distinct challenges and opportunities that emerge in this next frontier of generative collaboration.}

\subsection{Future research direction}
Building on our findings and research goals, we identify several key directions for advancing the understanding of human-AI collaboration through a CSCW lens, particularly in exploring agent-to-agent interactions, human agency and reliance on AI, and a human-centered toolkit development for improving the collaborative outcome.

Our findings relate to the challenges of managing collaboration across multiple agents where issues with error propagation, looping behaviors, and unpredictability in agent-to-agent interactions were observed. This highlights a need to understand how independent AI agents coordinate and make decisions collectively, especially in scenarios where human oversight may be limited. This direction offers the potential to extend CSCW principles to new inter-agent communication paradigms, advancing collaboration frameworks that account for complex, multi-agent interactions.

Future research should investigate where human mediation can add value in multi-agent collaborations, particularly in high-stakes or complex environments. Understanding the optimal intervention points could help identify when and how users should exercise control, balancing the need for autonomous AI operations with the benefits of human oversight. This research direction would support the design of systems that reinforce human agency, ensuring that users can guide AI behaviors and maintain critical decision-making power. As reliance on AI systems grows, so does the need to mitigate risks of over-reliance or excessive skepticism \cite{passi_appropriate_2024}. Future research could explore strategies to cultivate critical thinking in users, empowering them to discern when to trust or question AI outputs.

Based on the user feedback, there is an opportunity we plan to pursue to develop a {\itshape Multi-Agent GenAI UX Toolkit} through an iterative co-design approach that supports effective collaboration between humans and AI agents. \textcolor{black}{The idea for this toolkit emerged from our prior design-oriented research on early adopters' workflows and tool development \cite{naik_designing_2025}; building on that foundation, the current study contributes additional empirical insights, particularly around modes of collaboration, division of agency/control, facilitating transparency, and agent role-customization as collaborator, that further shape the toolkit to support more accessible and effective collaborative workflows with multi-agent generative AI.}

Through these avenues, future research can expand the CSCW literature on multi-agent generative AI systems, enhancing our understanding of new collaboration dynamics, design paradigms for the nuances of agent-to-agent collaboration and human-agent collaboration, and the development of robust, human-centered toolkit.

\section{Conclusion}
This study reveals how early adopters of multi-agent generative AI systems conceptualize these tools as collaborative partners, attributing distinct roles and expectations to each AI agent within a collaborative workflow. Early adopters appreciated the multi-agent Gen AI's capabilities, such as task specialization, workflow automation, divide and conquer, and role-based or task-based human-like conversable agents. Our findings identify \textcolor{black}{ collaboration patterns specific to multi-agent generative AI}: AI-dominant characteristics, which enhance efficiency yet present challenges of unpredictability, and AI-assisted characteristics, which prioritize structured and reliable interactions, often preferred in high-stakes environments. However, significant challenges, including unpredictability, error propagation, unproductive loops, and lack of transparency, were also identified which hindered the collaboration. Transparency emerges as a critical factor in both models, with layered transparency mechanisms fostering trust and enabling user control across varying levels of expertise. The study uncovered perceptions around transparency, with developers requiring detailed visibility for debugging and preventing biases, while prioritizing simplicity to build trust. Additionally, early adopters revealed the importance of adaptable, role-based or task-based AI agents to achieve a shared goal, that integrate seamlessly into existing workflows, aligning with the specific needs of diverse users. These insights underscore the necessity of transparency, customization, and role differentiation in the design of human-centered, multi-agent AI systems, contributing to the advancement of human-AI collaboration within CSCW research.

The limitations of this study include the small sample size and its focus on early adopters of one particular technology organization, which may not represent the broader early adopter population. Furthermore, the participants' experiences were limited to prototype systems or internal team use only, leaving open questions about real-world scalability and integration. The study helps identifying more opportunity areas for future research for deeper understanding of these human-to-agent and agent-to-agent interactions, without making the humans blindly rely on the AI outputs but have a critical thinking approach. These areas include the need to explore transparency mechanisms, collaborative mechanisms based on AI tool's capabilities, communication strategies between humans and agents to inform the design of future multi-agent Gen AI systems of specialized agents for seamless collaboration.

\input{main.bbl}

\end{document}